\documentstyle[11pt]{article}

\begin{document}

\vspace*{0.5cm}

\begin{center}
\large {\bf Laser Pulses as Measurements. Application to the
Quantum Zeno Effect} \\

\vspace*{.5cm}
   Almut Beige\footnote{e-mail: beige@theorie.physik.uni-goettingen.de}, 
   Gerhard C. Hegerfeldt\footnote{e-mail: hegerf@theorie.physik.uni-
   goettingen.de} 
   and Dirk G. Sondermann\footnote{e-mail: 
   sonderma@theorie.physik.uni-goettingen.de} \\ [.5cm]
\normalsize
   Institut f\"ur Theoretische Physik\\
   Universit\"at G\"ottingen\\
   Bunsenstr. 9\\
   D-37073 G\"ottingen, Germany
\end{center}

\begin{abstract} Short pulses of a probe laser have been used in the
past to measure whether a two-level atom is in its ground
or excited state. The probe pulse couples the ground state to a
third, auxiliary, level of the atom. Occurrence or absence of
resonance fluorescence were taken to mean that the atom was found in its
ground or excited state, respectively. In this paper we investigate
to what extent this procedure results in an effective measurement to
which the projection postulate can be applied, at least
approximately. We discuss in detail the complications arising from
an additional time development of the two-level system proper during 
a probe pulse. We extend our previous results for weak probe pulses
to the general case and show that one can model an ideal 
(projection-postulate) measurement  much better with a strong than a
weak probe pulse. 
In an application to the quantum Zeno effect we calculate the
slow-down of the atomic time development under $n$ repeated probe
pulse measurements and determine the corrections compared 
to the case of $n$ ideal measurements. \\[.25cm]
PACS numbers  42.50.-p; 32.80.-t; 03.65.Bz
\end{abstract}

\vspace*{0.5cm}

\noindent {\bf 1. Introduction}

\vspace*{0.5cm}

\noindent 

For a quantum system a measurement in general involves a macroscopic
apparatus which interacts with the system and leaves a pointer in
some definite position. 
By an {\em ideal} measurement at time $t$ we mean a measurement whose
effect on a state can be described by applying the projection
postulate at time $t$. 
The  projection postulate as commonly used nowadays is due to 
L\"uders \cite{Lue}.
For observables with degenerate eigenvalues his formulation 
differs from that of von Neumann \cite{Neu}. The projection postulate
has been widely  regarded as a useful tool\footnote{%
L\"uders stressed its provisional character:
 ``The projection postulate will be employed only until a better
   understanding of 
   the actual measurement process has been found'' 
(G. L\"uders, private communication to G.C.H.).}.  
It is also known that one can envisage more general measurements where the 
projection postulate is not applicable; cf. e.g. \cite{Lud}.

Some time ago, Cook \cite{Cook}
proposed to measure the ground or excited state (stable or
metastable) of two-level a system 
by means of a short pulse of a probe laser
which pumps the transition between the ground state, level 1,
and an auxiliary rapidly decaying third level  (see Fig.~1). After
emission of a photon the atom will be in its ground state, and so it
is natural to assume that occurrence of resonance fluorescence means
that the atom is in level 1 at the end of the pulse.
Absence of resonance fluorescence
was assumed  to imply that the atom is in level 2.
This idea was subsequently 
used in an experiment of Itano et al.~\cite{Wine} to
test the so-called quantum Zeno effect (QZE) \cite{MiSu}.

In a recent paper \cite{BeHe} two of the present authors have
investigated to what extent a short probe pulse  as in Refs.~\cite{Cook,Wine} 
can be regarded as an effective measurement to which the 
projection postulate can be applied, at least approximately.  We
distinguished two cases. If there is no interaction between levels 1 and
2  during a probe
pulse, we showed that a single probe pulse does indeed project on either 
$|1 \rangle$ or $|2 \rangle$, up
to terms decreasing exponentially with its duration. If one has  an
additional small driving field (Rabi frequency $\Omega_2$) 
between levels 1 and 2 during a probe pulse the situation is more
complicated, and we had to restrict ourselves to {\em weak} probe
pulses (with Rabi frequency $\Omega_3$ much smaller than the Einstein
coefficient $A_3$ of level 3, but still $\Omega_2 \ll \Omega_3$). 
Under this additional assumption we were able to  show that an
atom {\em with} photon emission during the probe pulse is projected
onto a state (density matrix)  extremely close, but not
identical to, $|1 \rangle \langle 1|$, and that an atom {\em without}
photon emission is projected onto a state  very close
to $|2 \rangle \langle 2|$. The small difference to the ideal
projection-postulate result was explicitly calculated
to first order in the small parameters
\begin{eqnarray}\label{eps}
\epsilon_{\rm p} \equiv \frac{\Omega_2 A_3}{\Omega_3^2},~~~
\epsilon_R \equiv \frac{\Omega_2}{\Omega_3}
 ~~&{\rm and}&~~ 
\epsilon_A \equiv \frac{\Omega_2}{A_3} ~.
\end{eqnarray}
We also pointed out that these $\epsilon$
parameters had to be much smaller than 1 for a pulse of the probe
laser to act as an effective measurement. For many probe pulses
(``measurements'') this small difference adds up. The cumulative effect was
explicitly determined, and  we used our results to analyze the
experiment of Ref.\ \cite{Wine} on the QZE.

The QZE  \cite{MiSu} deals with rapidly repeated measurements on 
a system. Under the assumptions of $n$ ideal measurements,
at times $\Delta t, \dots ,$ $ n\Delta t$ $ = t$, $t$ fixed,
the QZE predicts a 
slow-down of the time development and 
ultimately a freezing of the state for $\Delta t \rightarrow 0$, or 
$n \rightarrow \infty$. The QZE has
found a tremendous interest in the literature \cite{interest}.

In the experiment of Ref.\ \cite{Wine} several thousand ions are stored
in a trap and an $rf$ field in resonance is used to drive the transition
between two levels 1 and 2, which are essentially stable. The $rf$
field is a so-called $\pi$ pulse of duration $T_\pi$ which changes
state $|1 \rangle$ into state $|2 \rangle$. The experiment then
endeavors to perform $n$ measurements of the populations of levels 1
and 2 between $0$ and $T_\pi$ by means of $n$  short pulses of a
probe laser (see Fig.~2) which pumps the transition between level 1 
and the auxiliary rapidly decaying level 3, with occurrence of 
resonant fluorescence from an atom  taken to mean detection of the 
atom in level 1 and absence thereof as detection 
of the atom in level 2, as proposed in Ref.\ \cite{Cook}.
At time $T_\pi$ the actual population of levels 1 and 2 are then
determined through the resonant fluorescence of all atoms in the trap
under the influence of a long probe pulse, with the $\pi$ pulse
switched off. The  populations determined in this way 
are in good agreement with the predictions of the QZE .

The relevance of this experiment for the QZE has been contested in 
the literature. It was pointed out by some authors 
\cite{crit1,Block,Schenzle,crit2,crit3,crit4,crit5} that there is no
need to use the QZE to explain the experiment. One could simply 
incorporate the auxiliary level and the driving by the probe laser 
together with the $rf$ field in the 
Hamiltonian or in the  Bloch equations of the corresponding
three-level system.  From a 
numerical solution of the Bloch equations one does indeed obtain
the final population at time $T_\pi$ in good agreement with the
experiment \cite{Schenzle}. The probe pulses are, in this view, 
not measurements but part of the dynamics.

In our previous paper \cite{BeHe} we used our results on the close,
but not perfect,
resemblance of individual probe pulses  with ideal measurements 
to determine in closed form, to first order in the small $\epsilon$
parameters, the population of level 2 after $n$ probe
pulses, at the end of the $\pi$ pulse. 
This was compared with a {\em numerical} solution of the
three-level Bloch equations, and an amazing agreement was found. In
Ref.\ \cite{BeHe} our conclusion was that the projection postulate is a
useful tool to give a quick and approximate understanding of the type
of experiment performed in Ref.\ \cite{Wine}, but that for a more precise
description one needs a more detailed analysis of the actual
measurements, as for example carried out by us in Ref.\ \cite{BeHe}.

In the present paper we treat the general case of arbitrary, not
necessarily weak,  probe pulses.
The duration of a probe pulse, $\tau_{\rm p}$, is assumed to be less 
than $T_{\pi}= \pi /\Omega_2$, and   
our results will be given to first order in the above parameters
$\epsilon_{\rm p}$, $\epsilon_R$ and $\epsilon_A$.
 It will turn out that 
with increasing $\Omega_3$, i.e.\ increasing strength,
the probe pulse models an ideal measurement 
with projection postulate much better than for weak pulses; this is 
particularly true in the presence of the $rf$ field. 

The plan of the paper is as follows. In Section 2 we deal with a
single short probe pulse during which the $rf$ field is also switched
on, and the description of individual atoms is studied according to
whether they do or do not emit photons during a probe pulse. We show
that at the end of a probe pulse an atom can only be in one of two
states, denoted\footnote{%
A tilde on a density matrix indicates that its trace
is $1$. We will also have occasion to use non-normalized density
matrices. 
Here, a state can be pure or mixed.}
by  $\tilde{\rho}^>$ and $\tilde{\rho}^0$, respectively, 
depending on whether or not the atom has 
emitted photons. These states still contain a contribution
 from level 3, which will decay during a short 
transient time. During this decay 
time the $rf$ field will also change the state, and the state will
depend on the particular decay time chosen. We show, however, that
there are two uniquely defined states 
which effectively describe the resultant action at the
end of a probe pulse and which one can regard as the projected
states of an atom. The two states are independent of the initial
state, have no contribution
 from level 3 and are independent of any particular choice of transient
decay time,
These uniquely defined states are, however, never
exactly realized by the atom  and are in this sense virtual and a
mathematical tool.

In Section 3 we consider repeated probe pulses in connection with the
QZE and the cumulative effect of the corrections to the projection
postulate. Closed expressions are given and compared to numerical
solutions of the three-level Bloch equations. Our conclusion is that
if one would perform the experiment of Ref.\ \cite{Wine} with a much
stronger probe pulse (with Rabi frequency $\Omega_3$ of the order of 
the Einstein coefficient $A_3$) then the result for the level population would
be much closer to the case of ideal projection-postulate 
measurements  than for weak probe pulses, as in the original experiment
of Ref.\ \cite{Wine}.

In Section 4 we discuss our results, and in the Appendix we 
briefly review  the quantum jump approach \cite{HeWi,Wi,He,HeSo} which we
use in this paper. The quantum jump approach
is essentially equivalent to quantum trajectories \cite{QT} and
to the Monte-Carlo wave function approach \cite{MC}. 

\vspace*{0.5cm}

\noindent {\bf 2. Effect of a single probe pulse on the atomic state}

\vspace*{0.5cm}

In this Section we discuss how a single probe pulse acts as 
an effective measurement of an atomic level and to what extent
the usual projection postulate can be applied.

The case where the $rf$ field is switched 
off while the probe pulse is on was discussed in some detail
in Ref.\ \cite{BeHe}. It was shown that if the length of the probe pulse
$\tau_{\rm p}$ satisfies the condition 
\begin{eqnarray} \label{cond}
\tau_{\rm p} & \gg & {\rm max}\left\{ A_3^{-1},~A_3/\Omega_3^2 \right\}~,
\end{eqnarray}
then the probe pulse provides an effective 
reduction of the initial state to $|1 \rangle$ or $|2 \rangle$ 
depending whether or not  photons were emitted (allowing, in the
former case, for a decay of the third-level contributions). The
respective probabilities are the same
as predicted by the 
projection postulate  for an ideal measurement of the level
populations.  

If the $rf$ field is switched on during the probe pulse 
it causes a small pumping between level 1 and 2. 
However, we will show that, independent of its state at the beginning
of a probe pulse, right at the end of the pulse an atom will be in
one of two states, $\tilde{\rho}^0$ and $\tilde{\rho}^>$, depending
on whether or not it has emitted photons. To allow for the decay of
the third-level contributions a transient time at least of the order of
several decay times $A^{-1}_3$ is needed. During this time the $rf$
field is also active, and therefore the result has a small dependence on
the transient time chosen. At the end end of this section we
therefore mathematically construct two density matrices,
$\tilde{\rho}^0_{\rm P}$ and $\tilde{\rho}^>_{\rm P}$, in the 1-2
subspace which are independent of the transient time. These states,
although not physically realized, can be used in a consistent way as
hypothetical or virtual states of an atom at the end of the probe
pulse, i.e.\ as states on which the probe pulse projects. 

\vspace*{0.5cm}

\noindent 2.1 {\em Subensemble} without {\em photon emission} 

\vspace*{0.5cm}

\noindent We now determine the state at the end of the probe pulse
for an atom
which has not emitted photons. The result is the same as that found in
our
Ref.\ \cite{BeHe}, but the present derivation is more streamlined.
According to Eq.~(\ref{2.15}) of the Appendix, an  atom
without photon emission evolves with \cite{HeWi,He,HeSo}
\begin{equation}\label{i1}
U^I_{{\rm red}} (t,0) =
\exp \left\{ - {\rm i} H^I_{{\rm red}} t/ \hbar \right\}
\end{equation}
where the reduced (or conditional) Hamiltonian is given by
Eq.~(\ref{2.12}) of the Appendix,
\begin{equation}\label{i2}
{\rm i} H^I_{{\rm red}} / \hbar \equiv M = M_0 + M_1
\end{equation}
with
\begin{eqnarray}\label{i3}
M_0 &=& \frac{{\rm i}}{2}~\Omega_3 \left\{ | 3 \rangle \langle 1 | + | 1
\rangle \langle 3| \right\} + ~\frac{1}{2}~A_3 |3 \rangle \langle 3| \\
\label{i4}
M_1 &=& \frac{{\rm i}}{2}~ \Omega_2 \{| 2 \rangle \langle 1 | + | 1 \rangle
\langle  2 | \}~.
\end{eqnarray}
The eigenvalues of $M_0$ are $\lambda^0_2=0$ and
\begin{equation}\label{i5}
\lambda^0_{1,3} = ~\frac{1}{4}~ A_3 \left\{ 1 \pm \sqrt{1 - 4
\Omega^2_3 / A^2_3} \right\}~.
\end{equation}

For $M = M_0 + M_1$ these go over into $\lambda_i$, $i = 1,2,3$, which
are the roots of the cubic equation
\begin{equation}\label{i6}
p (\lambda) \equiv {\rm det} (M-\lambda) 
=\lambda^3-\frac{1}{2}A_3\lambda^2+\frac{1}{4}(\Omega_2^2+\Omega_3^2)
\lambda-\frac{1}{8}A_3\Omega_2^2
= 0~.
\end{equation}
Newton's method or a linear expansion gives\footnote{%
In our case $\lambda_2$ is real and one has the explicit bounds
$$
\frac{1}{2}~\Omega_2 \epsilon_{\rm p} (1 + \epsilon_R^2)^{-1} \le \lambda_2
 \le ~\frac{1}{2}~\Omega_2 \epsilon_{\rm p} (1 + \epsilon^2_{\rm p})(1 +
\epsilon^2_R)^{-1}
$$
since $p(\lambda)$ changes sign between these bounds, as seen by insertion.}
\begin{equation}\label{i7}
\lambda_2 = ~\frac{1}{2}~\Omega_2 \epsilon_{\rm p} ( 1 + {\cal O}
({\mbox{\boldmath $\epsilon$}}^2 ))
\end{equation}
where ${\cal O}({\mbox{\boldmath $\epsilon$}}^2)$ denotes a correction at 
least quadratic in ${\mbox{\boldmath $\epsilon$}} \equiv (\epsilon_{\rm p}, 
\epsilon_R)$, so that ${\cal O}({\mbox{\boldmath $\epsilon$}}^2) / 
| {\mbox{\boldmath $\epsilon$}} |^2$ remains bounded. For $\lambda_{1,3}$, 
one obtains
\begin{equation}\label{i8}
\lambda_{1,3} = \lambda^0_{1,3} (1 +{\cal O}({\mbox{\boldmath $\epsilon$}}))~.
\end{equation}
If the $\lambda_i$ are different (which here means
$\lambda_1 \not=\lambda_3$) then $M$ has three normalized
eigenvectors $| \lambda_i \rangle$,  which are in general not
orthogonal. The reciprocal basis vectors $| \lambda^i
\rangle$ are defined by
\begin{equation}\label{i9}
\langle \lambda_i | \lambda^j \rangle = \delta_{ij}
\end{equation}
and satisfy the eigenvalue equation
\begin{equation}\label{i10}
M^\dagger |\lambda^i \rangle = \bar{\lambda}_i | \lambda^i \rangle~.
\end{equation}
In particular, an elementary calculation gives
\begin{eqnarray} \label{i11}
| \lambda_2 \rangle &=& - {\rm i} \epsilon_{\rm p} |1 \rangle + | 2 \rangle -
\epsilon_R | 3 \rangle + {\cal O} ({\mbox{\boldmath $\epsilon$}}^2)~ 
\\ \label{i12}
| \lambda^2 \rangle &=& {\rm i} \epsilon_{\rm p} | 1 \rangle + | 2 \rangle -
\epsilon_R | 3 \rangle + {\cal O} ({\mbox{\boldmath $\epsilon$}}^2)
\end{eqnarray}

For ${\rm e}^{-M t}$ one has the representation\footnote{%
Alternatively,
$$
{\rm e}^{-M t} = \frac{(M-\lambda_2)(M-\lambda_3)}
{(\lambda_1-\lambda_2)(\lambda_1-\lambda_3)} {\rm e}^{-\lambda_1t}
+{\rm ~cyclic~permutations~,}
$$ 
which is directly checked by application to the eigen\-vec\-tors. 
Comparison with Eq.~(\ref{i13}) gives explicit
expressions for $|\lambda_i \rangle \langle \lambda^i|$. In the above
formula one can take the limit $\lambda_3 \rightarrow \lambda_1$. This
leads to a term with the factor $t \exp\{-\lambda_1 t\}$. The following
arguments are then easily adapted to this degenerate case.}
\begin{equation}\label{i13}
{\rm e}^{-M t} = \sum_{i=1}^3 {\rm e}^{-\lambda_i t} 
|\lambda_i \rangle \langle \lambda^i|
= \sum_{i=1}^3 {\rm e}^{- \lambda_i t} P_i
\end{equation}
with $P_i \equiv| \lambda_i \rangle \langle \lambda^i |$. 
If $\tau_{\rm p}$ satisfies the condition of Eq.~(\ref{cond}) then
the ex\-po\-nen\-tials $\exp\{-\lambda_{1,3} \tau_{\rm p}\}$ can be 
neglected, while $\exp\{-\lambda_2 \tau_{\rm p}\} \approx 1$. Therefore for 
any initial pure state $\left|\psi\right>$ one has
\begin{equation}\label{i14}
{\rm e}^{-M \tau_{\rm p}} | \psi \rangle 
= {\rm e}^{-\lambda_2 \tau_{\rm p}} \langle \lambda^2 |
\psi \rangle | \lambda_2 \rangle~ \mbox{+ exponentially small terms}
\end{equation}
which is proportional to $| \lambda_2 \rangle$. Thus at the end of a
probe pulse the subensemble without photon emissions is, after
normalization and up to
exponentially small terms, in the state $| \lambda_2 \rangle$ and thus
independent of the initial state.

The probability $P_0 (\tau_{\rm p}; |\psi \rangle)$ for no emission
during a probe pulse, for initial state $|\psi \rangle$ is the
norm-squared of Eq.~(\ref{i14}). If one has a density matrix 
$\rho =\sum_i \alpha_i |\psi_i \rangle \langle \psi_i |$
at the beginning of a probe pulse the no-emissions probability
becomes
\begin{equation}
P_0 (\tau_{\rm p};\rho) =\sum_{i} \alpha_i P_0(\tau_{\rm p};|\psi_i\rangle)~.
\end{equation}
By Eqs.~(\ref{i11}) and (\ref{i12}) one obtains
\begin{equation}\label{i16}
P_0 (\tau_{\rm p} ; \rho)
=\rho_{22}-\epsilon_{\rm p} \pi \frac{\tau_{\rm p}}{T_\pi}
\rho_{22}+2\epsilon_{\rm p} \mathop{\rm Im}\rho_{12}-
2\epsilon_R \mathop{\rm Re}\rho_{23} 
+ {\cal O} ({\mbox{\boldmath $\epsilon$}}^2)~. 
\end{equation}
We thus arrive at 
\begin{eqnarray}\label{i17}
\rho^0 (\tau_{\rm p};\rho) 
& \equiv & {\rm e}^{-M \tau_{\rm p}}
 ~\rho~ {\rm e}^{-M^\dagger \tau_{\rm p}}\nonumber
 ~ = ~ P_0 (\tau_{\rm p} ; \rho) |\lambda_2 \rangle \langle \lambda_2|
\end{eqnarray}
and, for the basis $|1 \rangle$, $|2 \rangle$, $|3 \rangle$, we
obtain from Eq.~(\ref{i11})
\begin{eqnarray}\label{i17a}
\tilde \rho^0 & \equiv & |\lambda_2 \rangle \langle \lambda_2|
 ~ = ~ \left( \begin{array}{ccc} 
0 & -{\rm i} \epsilon_{\rm p} & 0 \\
{\rm i} \epsilon_{\rm p} & 1 & -\epsilon_R \\
0 & -\epsilon_R & 0 \end{array} \right)
+{\cal O} ({\mbox{\boldmath $\epsilon$}}^2)~. 
\end{eqnarray}

We have thus shown that, up to exponentially small terms, a probe
pulse projects each atom with no photon emission onto the state 
$|\lambda_2 \rangle$, which is  
close to $|2 \rangle$. The probability for this is given 
by $P_0 (\tau_{\rm p}; \rho)$ of Eq.~(\ref{i16}) where $\rho$ is the density 
matrix of the atom at the beginning of the probe pulse. In case the
$rf$ field is switched off $(\Omega_2=0)$ during the probe pulse,
one has $|\lambda_2 \rangle =|2 \rangle$ and the probe pulse projects
onto $|2 \rangle$ with probability $\rho_{22}$, up to exponentially small 
terms.

\vspace*{0.5cm}

\noindent 2.2 {\em Subensemble} with {\em photon emissions}

\vspace*{0.5cm}

If an atom does emit photons during a probe pulse, the emissions 
can occur at 
random times. Therefore, even if one had started with a pure state 
$| \psi \rangle$ the subensemble of all atoms with emissions is described 
by a mixture. The corresponding density matrix is denoted by 
$\rho^>(\tau, \rho)$, with $\rho$ the  density matrix at the beginning of
a probe pulse, and the normalization of $\rho^>$ is chosen such that 
$\mathop{\rm tr} \rho^>(\tau, \rho)$ is the probability for emissions until 
time $\tau$. The aim of this subsection is to show that 
$\rho^> (\tau_{\rm p}, \rho)$
is in general proportional to a $\rho$-independent 
matrix. Hence the subensemble with photon emissions is, at the end of a 
probe pulse, described by a fixed normalized density matrix 
$\tilde{\rho}^> \equiv \rho^> / \mathop{\rm tr} \rho^>$, which is 
independent of the initial state. This density matrix
$\tilde{\rho}^>$, which still contains contributions of level 3, will
now be explicitly determined.

We denote by $I(\tau'; \rho)$ the probability density for the emission
of a photon at time $\tau'$. Now, after a particular atom has emitted its 
last photon before $\tau$, at time $\tau'$ say, the atom is in its 
ground state $|1\rangle$ and until $\tau$ the time development is given by 
\begin{equation}\label{H1}
{\rm e}^{-M(\tau-\tau')} |1 \rangle \langle 1|
{\rm e}^{-M^\dagger(\tau-\tau')}
\equiv \rho^0(\tau-\tau';|1 \rangle)
\end{equation}
Hence the subensemble is described by 
\begin{eqnarray}\label{H2}
\rho^>(\tau; \rho) & = & \int^{\tau}_0 {\rm d}\tau' I(\tau'; \rho)
\rho^0(\tau - \tau'; |1\rangle), \nonumber\\
& = & \int^{\tau}_0 {\rm d}\tau' I(\tau - \tau' ; \rho)
\rho^0 (\tau';|1\rangle)~.
\end{eqnarray}
This is also seen directly from Eq.~(\ref{2.B3}) of the Appendix. 
The density matrix of the 
complete ensemble is given by
\begin{equation}\label{H2c}
\rho(\tau ;\rho) = \rho^0(\tau ;\rho) + \rho^>(\tau ;\rho),
\end{equation}
and 
\begin{equation}\label{H2b}
\mathop{\rm tr} \rho^>(\tau; \rho) 
= 1 - \mathop{\rm tr} \rho^0 (\tau; \rho)
= 1 - P_0 (\tau; \rho)~.
\end{equation}
Since in our case 
$I (\tau ;\rho) = A_3 \rho_{33}(\tau ;\rho)$
by Eq.~(\ref{2.16d}) of the Appendix, Eqs.~(\ref{H2}) and 
(\ref{H2c}) give the integral equation 
\begin{equation}\label{He3}
I(\tau; \rho) = A_3 \rho^0_{33} (\tau ;\rho) + A_3~ \int^\tau_0 {\rm d}
\tau' I(\tau - \tau' ;\rho) \rho^0_{33}  (\tau';|1\rangle).
\end{equation}
This can be solved by Laplace transform\footnote{%
Alternatively one could solve the Bloch equations for $\rho$.}.
Since, by Eq.~(\ref{H1}), $\rho^0_{33}$ is a sum of exponential
terms whose exponents do not depend on $\rho$, $I$ is of the form 
\begin{equation}\label{H4}
I(\tau ;\rho) = c_0 + c_1 {\rm e}^{-\mu_1 \tau} +~\sum^9_{\alpha = 2}
c_\alpha {\rm e}^{- \mu_\alpha \tau}
\end{equation}
where $c_i=c_i(\rho)$. One has $\mathop{\rm Re} \mu_i > 0$, $\mu_1$ is 
real and of the order of $\lambda_2$, while all the other $\mu_\alpha$'s are 
of the order of $\lambda_1$ and $\lambda_3$ \footnote{%
For $\Omega_2 = 0$ one has $\mu_1 = 0$, and in general 
$\mu_1$ is quadratic in 
$\Omega_2$. The $\mu_i$'s are in fact the eigenvalues of the matrix 
in the Bloch equations. Two of the eigenvalues vanish for $\Omega_2 = 0$. 
 From Eq.~(\ref{He3}) one can determine $\mu_1$ explicitly as 
$\mu_1 = 2 \epsilon_{\rm p} \Omega_2 (A^2_3 + \Omega^2_3) / (A^2_3 + 2 
\Omega^2_3) + {\cal O} (\Omega^4_2)$.}. 

Therefore one sees that $c_0$ is the stationary emission rate of the
three-level system.
Furthermore, considering times of the order of the length of the 
probe pulse, i.e.\ $\exp\{-\mu_1 \tau\} \approx 1$ and 
$\mathop{\rm Re} \mu_\alpha \tau 
\gg 1$ for $\alpha \ge 2$, one sees that $c_0 + c_1$ must be positive. 
Physically the emission rate for such times cannot exceed 
the stationary emission rate $A_3\Omega_3^2/(A_3^2+2\Omega_3^2)$
of the 1-3 system with $\Omega_2 = 0$
and initial state $| 1 \rangle$, and the same is true for very large
$\tau$, for which $\exp\{-\mu_1 \tau\} \ll 1$. 
Hence one has the inequalities
\begin{equation}\label{H5}
0 \le \left\{ \begin{array}{c} c_0 \\ c_0 + c_1(\rho) \end{array}
\right\} 
\le ~\frac{A_3 \Omega^2_3}{A^2_3 + 2 \Omega^2_3}~.
\end{equation}
Therefrom one obtains
\begin{equation}\label{H6}
c_0,~ |c_1 (\rho)| ~\le ~\frac{A_3 \Omega^2_3}{A^2_3 + 2 \Omega^2_3}~.
\end{equation}
Only these rough inequalities will be needed\footnote{%
 From Eq.~(\ref{He3}) one obtains by a more detailed calculation
\begin{eqnarray}
c_0 & = &  \frac{1}{2} \frac{A_3 \Omega_3^2}{A_3^2+\Omega_3^2} \nonumber\\
c_1 (\rho) & = &  c_0 \left( \frac{A_3^2}{A_3^2+2\Omega_3^2}
-2\frac{A_3^2+ \Omega_3^2}{A_3^2+2\Omega_3^2} \langle
\lambda_2 |\rho| \lambda_2 \rangle \right)~.    \nonumber
\end{eqnarray}}.

 From Eq.~(\ref{i13}) for ${\rm e}^{-Mt}$ one obtains
\begin{equation}\label{H8}
\rho^0 (\tau ; |1\rangle) = ~\sum_{i,j} P_i |1 \rangle \langle 1| P_j^\dagger
{\rm e}^{-(\lambda_i + \bar{\lambda}_j) \tau }
\end{equation}
where $P_i$ was defined as $|\lambda_i \rangle \langle \lambda^i|$.
Inserting Eqs.~(\ref{H4}) and (\ref{H8}) into Eq.~(\ref{H2}) with
$\tau = \tau_{\rm p}$ gives 
\begin{eqnarray}\label{H9}
\rho^> (\tau_{\rm p} ; \rho) & = &\left(
 c_0 + c_1 {\rm e}^{-\mu_1 \tau_{\rm p}} \right)
 ~\int^{\tau_{\rm p}}_0 {\rm d}\tau' \rho^0 (\tau';|1\rangle) \nonumber\\
& & + c_1 ~ \int^{\tau_{\rm p}}_0 {\rm d}\tau'
\left( {\rm e}^{-\mu_1 (\tau_{\rm p}-\tau')}-
{\rm e}^{-\mu_1 \tau_{\rm p}} \right)
\rho^0 (\tau' ; |1\rangle) \nonumber\\
& & + \int^{\tau_{\rm p}}_0 {\rm d}\tau' ~\sum_{\alpha \ge 2} c_\alpha
{\rm e}^{-\mu_\alpha (\tau_{\rm p}- \tau')} {\rm e}^{-2 \lambda_2 \tau' }
P_2 |1 \rangle \langle 1| P_2^\dagger \nonumber \\
& &+ \int^{\tau_{\rm p}}_0 {\rm d}\tau' ~\sum_{\alpha \ge 2} c_\alpha
{\rm e}^{-\mu_\alpha (\tau_{\rm p} - \tau')} \sum_{(i,j) \not= (2,2)}
{\rm e}^{-(\lambda_i + \bar{\lambda}_j) \tau'}
P_i | 1 \rangle \langle 1 | P_j^\dagger ~.
\end{eqnarray}
It will be shown that the third and the last term are of order 
${\mbox{\boldmath $\epsilon$}}^2$. The last term is clearly proportional to 
exponentially small terms and can therefore be omitted. Since 
$ {\rm e}^{-\mu_1 (\tau_{\rm p}-\tau')}-
{\rm e}^{-\mu_1 \tau_{\rm p}} \le \mu_1 
\tau'$, the second term is bounded by
\begin{eqnarray}\label{H10}
c_1 \mu_1 ~\int^{\tau_{\rm p}}_0 {\rm d}\tau' ~\tau' \rho^0 (\tau';|1\rangle)
&=&  c_1 \mu_1 ~\int^{\tau_{\rm p}}_0 {\rm d}\tau' ~\tau'
 {\rm e}^{-2 \lambda_2 \tau'}
P_2 |1 \rangle \langle 1| P^\dagger_2 \nonumber\\
& &+ c_1 \mu_1 ~\sum_{(i,j) \not= (2,2)} 
\int^{\tau_{\rm p}}_0 {\rm d}\tau' ~\tau'
{\rm e}^{-(\lambda_i + \bar{\lambda}_j) \tau'}
P_i |1 \rangle  \langle 1| P^\dagger_j~.
\end{eqnarray}
Now, $P_2 | 1 \rangle = | \lambda_2 \rangle \langle \lambda^2 | 1
\rangle$ is of order ${\mbox{\boldmath $\epsilon$}}$, by Eq.~(\ref{i12}), 
and $P_2 | 1 \rangle \langle 1| P_2^\dagger$ is thus of 
order\footnote{\label{f}
Normalization of $\rho^>$ involves division by ${\rm tr} \rho^> =
1 - P_0(\tau_{\rm p}, 
\rho)$. For $\rho$ close to
$| \lambda_2 \rangle \langle\lambda_2| $ this is of order 
${\mbox{\boldmath $\epsilon$}}$, and in this case Eq.~(\ref{H11b}) holds 
only up to order $\epsilon_{\rm p} T_\pi /\tau_{\rm p}$.
In the subsequent applications, 
however, this will play no role.}
${\mbox{\boldmath $\epsilon$}}^2$.
The second integral in
Eq.~(\ref{H10}) is proportional to $(\lambda_i + {\overline \lambda_j})^{-2}$. 
Multiplied by $c_1 \mu_1$ this becomes bounded by ${\mbox{\boldmath 
$\epsilon$}}^2$. Thus the second term in Eq.~(\ref{H9}) is bounded by 
${\mbox{\boldmath $\epsilon$}}^2$. For the third term the argument is again 
based on $P_2 | 1 \rangle \langle 1 | P_2^\dagger = {\cal O}
({\mbox{\boldmath $\epsilon$}}^2)$. Hence 
\begin{equation}\label{H11}
\rho^> (\tau_{\rm p};\rho) = \left( c_0 + c_1 {\rm e}^{-\mu_1 \tau_{\rm p}}
\right) ~\int^{\tau_{\rm p}}_0
{\rm d}\tau' \rho^0 (\tau';|1\rangle) + {\cal O}
({\mbox{\boldmath $\epsilon$}}^2) 
\end{equation}
and thus one has for the normalized density matrix 
\begin{equation}\label{H11b}
\tilde{\rho}^> (\tau_{\rm p};\rho) = \tilde{\rho}^> (\tau_{\rm p}) 
 ~\propto~ \int^{\tau_{\rm p}}_0 {\rm d}\tau'~ \rho^0
(\tau';|1\rangle)+{\cal O}({\mbox{\boldmath $\epsilon $}}^2) 
\end{equation}
which is, up to order ${\mbox{\boldmath $\epsilon$}}^2$, independent of 
the initial state $\rho$ (cf. footnote {\ref{f}}).

The integral can be computed in closed form by choosing for 
$\rho$ in Eqs.~(\ref{H2}) and (\ref{H2c}) the stationary state $\rho^{ss}$ 
of the three-level system driven by $\Omega_3$ and $\Omega_2$. Then the
emission rate
$I = A_3 \rho_{33}^{ss}$ is time independent, and one obtains 
\begin{equation}\label{H11c}
\rho^{ss} = {\rm e}^{-M \tau_{\rm p}} \rho^{ss} {\rm e}^{- M^\dagger
\tau_{\rm p}}
+ A_3 \rho^{ss}_{33} ~\int^{\tau_{\rm p}}_0 {\rm d}\tau' ~\rho^0
(\tau';|1\rangle)~.
\end{equation}
With Eq.~(\ref{H11b}) this yields 
\begin{equation}\label{H11d}
\tilde{\rho}^> (\tau_{\rm p}) = \left( \rho^{ss} -
{\rm e}^{-M \tau_{\rm p}} \rho^{ss}
{\rm e}^{-M^\dagger \tau_{\rm p}} \right) /\mathop{\rm tr} (\cdot)
+ {\cal O}({\mbox {\boldmath $\epsilon $}}^2)~.
\end{equation}
One can express $\rho^{ss}$ directly in terms of $M$ as\footnote{%
 From Eqs.~(\ref{i3}) and (\ref{i4}) one finds 
$(M - \frac{1}{2} A_3) |3 \rangle = \frac{{\rm i}}{2} \Omega_3 | 1 \rangle$ 
and $M + M^\dagger =A_3 | 3 \rangle \langle 3 |$. From this one readily obtains
\begin{eqnarray*}
& & \int^{\tau_{\rm p}}_0 d\tau' ~ {\rm e}^{-M \tau'} | 1 \rangle \langle 1 |
{\rm e}^{- M^\dagger \tau'} \\
& = & - \frac{4}{\Omega_3^2 A_3} \left( M - \frac{1}{2} A_3 \right) 
\int^{\tau_{\rm p}}_0 d\tau' ~\frac{{\rm d}}{{\rm d}\tau'} \left( {\rm e}^{-M
\tau'}
{\rm e}^{-M^\dagger \tau'} \right)
\left( M^\dagger - \frac{1}{2} A_3 \right) \\
& = & \frac{4}{\Omega_3^2 A_3} \left( M- \frac{1}{2}A_3 \right) \left(
{\bf 1} - {\rm e}^{- M \tau_{\rm p}}
{\rm e}^{- M^\dagger \tau_{\rm p}} \right) \left(
M^\dagger - \frac{1}{2} A_3 \right)~.
\end{eqnarray*}
Comparison with Eqs.~(\ref{H11c}) and (\ref{H11d}) gives (\ref{H11f}).}
\begin{equation}\label{H11f}
\rho^{ss} = (M - \frac{1}{2} A_3) (M^\dagger - \frac{1}{2} A_3) 
/ \mathop{\rm tr} (\cdot)~.
\end{equation}

One can understand Eq.~(\ref{H11d}) as follows. It takes much longer
than the time $\tau_{\rm p}$ to reach the stationary three-level state, and
therefore there is not enough time to build up an appreciable
population of level 2. With the second term one just subtracts the
excess population of this level from $\rho^{ss}$ because 
\begin{equation}\label{H11e}
{\rm e}^{- M \tau_{\rm p}} \rho^{ss} {\rm e}^{- M^\dagger \tau_{\rm p}}
= {\rm e}^{-2 \lambda_2 \tau_{\rm p}} \langle \lambda^2 | \rho^{ss} | \lambda^2
\rangle |\lambda_2 \rangle \langle \lambda_2 | + {\rm exponentially
 ~small~terms}~.
\end{equation}

 From Eqs.~(\ref{H11d})-(\ref{H11e}) one can now calculate in a
straightforward way the normalized density matrix $\tilde{\rho}^>
(\tau_{\rm p})$ at the end of a probe pulse for the subensemble {\em with}
photon emissions and obtains
\begin{equation}\label{H11g}
\tilde{\rho}^> =~\frac{1}{A^2_3 + 2 \Omega^2_3 + A_3^2 \epsilon_{\rm p}
\Omega_2 \tau_{\rm p}} \left( \begin{array}{ccc}
A^2_3 + \Omega^2_3 & {\rm i} \epsilon_{\rm p} A^2_3 & {\rm i} A_3 \Omega_3 \\
-{\rm i} \epsilon_{\rm p} A^2_3 & A_3^2 \epsilon_{\rm p}
\Omega_2 \tau_{\rm p} & 
\epsilon_R (A^2_3+\Omega_3^2) \\
-{\rm i} A_3 \Omega_3 & \epsilon_R (A^2_3+\Omega_3^2) & \Omega^2_3
\end{array} \right) 
 ~+ {\cal O}({\mbox{\boldmath $\epsilon $}}^2)~.
\end{equation}
It is noteworthy that this is just, except for the $\epsilon$ terms,
the two-level stationary state of the 1-3 system (with $\Omega_2=0$). 
In addition $\tilde{\rho}^>$ has a 22-component proportional to 
$\tau_{\rm p}$. This results from the possibility of macroscopic dark
periods for the V system under consideration \cite{dark} so 
that the emission of photons may stop before the probe pulse ends. 
Such atoms are then in a state close to $|2 \rangle$ \cite{HeWi} and
contribute to $\tilde{\rho}^>_{22}$.

The effect of the probe pulse is therefore that right at its end an
atom, which is initially described by $\rho$, is projected with
probability $1 - P_0(\tau_{\rm p} ; \rho)$ onto $\tilde{\rho}^> (\tau_{\rm p})$
and with probability $P_0(\tau_{\rm p} ; \rho)$ onto ${\tilde\rho}^0$. We
note that $\tilde{\rho}^>$ and ${\tilde\rho}^0$ still have a
33 component which is small for ${\tilde\rho}^0$ but can be 
appreciable for $\tilde{\rho}^>$ and whose decay will now be studied.

\vspace*{0.5cm}

\noindent 2.3 {\em Decay of the 33 component and the question of measurement
  by a probe pulse}

\vspace*{0.5cm}

At the end of a probe pulse one still has $\Omega_2 \not= 0$ while
$\Omega_3 = 0$. The time development of an arbitrary density matrix is
then governed by the corresponding Bloch equations with $\Omega_3 =
0$, leading to a rapid decay of all $i 3$ and $3 i$ components, $i =
1,2,3$. Simultaneously the 1-2 transition is weakly pumped by
$\Omega_2$. If  $\tilde{\rho}_{3 3}^>$ is very small the net effect
of this is the same as increasing $\tilde{\rho}^>_{1 1}$ by
$\tilde{\rho}^>_{3 3}$ and replacing all $\tilde{\rho}^>_{i 3}$ and
$\tilde{\rho}^>_{3 i}$ by $0$ right after the probe pulse. 
This was the case in Ref.\ \cite{BeHe}.  If, however,
$\tilde{\rho}^>_{3 3}$ is not small, then the pumping from 1 to 2
during the decay is not quite the same with this replacement. 

Without employing Bloch equations one can very easily determine 
the time development by means of Eqs.~(\ref{2.B1})
and (\ref{2.B3}). Between probe pulses 
the {\em reduced} time development operator is given by
$\exp \{ -M_b \tau \}$ with 
\begin{eqnarray*}
M_b &=& \frac{{\rm i}}{2}~\Omega_2 \left\{ | 2 \rangle \langle 1 | + | 1
\rangle \langle 2| \right\} + ~\frac{1}{2}~A_3 |3 \rangle \langle 3|~.
\end{eqnarray*}
Hence, with 
\begin{eqnarray} \label{H12}
U_{\pi} (\tau+\tau_0, \tau_0) \equiv \left( \begin{array}{ccc}
\cos \frac{1}{2} \Omega_2 \tau & - {\rm i} \sin \frac{1}{2} \Omega_2
\tau & 0 \\
- {\rm i} \sin \frac{1}{2} \Omega_2 \tau & \cos \frac{1}{2} \Omega_2
\tau & 0 \\
0 & 0 & 1 \end{array} \right) \equiv U_{\pi} (\tau)
\end{eqnarray}
and with the projector $P_{12} \equiv | 1 \rangle \langle 1 |+ |
2 \rangle \langle 2 |$ one has 
\begin{equation}\label{H13}
{\rm e}^{- M_b \tau} = U_\pi (\tau) P_{12} + {\rm e}^{- \frac{1}{2} A_3 \tau}
| 3 \rangle \langle 3 |
\end{equation}
where $U_\pi $ describes the 1-2 pumping by the $rf$ field and
the last term the rapid decay of level 3. Until time $\tau_p+\tau$
the density matrix $\tilde{\rho}^> (\tau_{\rm p})$ has then developed, by
Eqs.~(\ref{2.16b}), (\ref{2.15}) and (\ref{2.B3}) of the Appendix, 
or by the analog of Eq.~(\ref{H2}), to 
\begin{eqnarray}\label{H14}
\tilde{\rho}^> (\tau_{\rm p} + \tau) & = & 
{\rm e}^{-M_b \tau} \tilde{\rho}^> (\tau_{\rm p}) 
{\rm e}^{-M_b^\dagger \tau} \nonumber\\
& &+ \int^\tau_0 {\rm d}\tau' ~I(\tau';\tilde{\rho}^> (\tau_{\rm p}))
{\rm e}^{-M_b(\tau- \tau')} | 1 \rangle \langle 1 | 
{\rm e}^{-M^\dagger_b (\tau - \tau')}
\end{eqnarray}
where, from Eq.~(\ref{H13}), 
$$
I (\tau';\tilde{\rho}^> (\tau_{\rm p})) 
= A_3 {\rm e}^{-A_3 \tau'} \tilde{\rho}^>_{33} (\tau_{\rm p})~.
$$
An analogous equation holds for the subensemble $\tilde{\rho}^0(\tau_{\rm p}
+\tau)$ of Subsection  2.1. During the decay of the 33 component an additional
photon may be emitted. Eq.~(\ref{H14}) is easily evaluated for times
$\tau$ larger than a transient time $\tau_{\rm tr}$ for which 
$\exp \{ -\frac{1}{2} A_3 \tau_{\rm tr} \}$
is at most of the order of ${\mbox{\boldmath $\epsilon $}}^2$ and
thus can be neglected.  E.g.\ for 
\begin{equation} \label{tr}
      \tau_{\rm tr} \cong 15/A_3
\end{equation}
the exponential is less than $10^{-4}$.
Eq.~(\ref{H14}) becomes, for $\tau \ge \tau_{\rm tr}$, 
\begin{eqnarray}\label{H15}
\tilde{\rho}^> (\tau_{\rm p} + \tau) & = & U_\pi (\tau) 
\Big\{ P_{12} \tilde{\rho}^> (\tau_{\rm p}) P_{12} \nonumber\\
& & + \tilde{\rho}^>_{33}(\tau_{\rm p}) \int^\tau_0 {\rm d}\tau' ~A_3
{\rm e}^{-A_3 \tau'} U_\pi^\dagger (\tau') | 1 \rangle \langle 1 | U_\pi
(\tau') \Big\} U_\pi^\dagger (\tau) ~.
\end{eqnarray}
With Eq.~(\ref{H12}) for $U_\pi$ the integral is readily evaluated. Denoting 
the matrix in the brackets by $\tilde{\rho}^>_{\rm P}$, one obtains
\begin{equation}\label{H16}
\tilde{\rho}^>_{\rm P} = \left( \begin{array}{ccc}
\tilde{\rho}^>_{11} (\tau_{\rm p}) + \tilde{\rho}^>_{33} (\tau_{\rm p}) & 
\tilde{\rho}^>_{12} (\tau_{\rm p}) - \frac{{\rm i}}{2}
\tilde{\rho}^>_{33}(\tau_{\rm p}) 
\epsilon_A & 0 \\
\tilde{\rho}^>_{21} (\tau_{\rm p}) + \frac{{\rm i}}{2}
\tilde{\rho}^>_{33}(\tau_{\rm p}) 
\epsilon_A & \tilde{\rho}^>_{22} (\tau_{\rm p}) & 0 \\  0 & 0 & 0
\end{array} \right)~.
\end{equation}
Insertion of $\tilde{\rho}^> (\tau_{\rm p})$ leads to
\begin{equation}\label{H17}
\tilde{\rho}^>_{\rm P} = 
\frac{1}{A^2_3 + 2 \Omega^2_3 + A^2_3 \epsilon_{\rm p} \Omega_2 \tau_{\rm p}} 
\left( \begin{array}{ccc} A^2_3 + 2 \Omega^2_3 
& {\rm i} \epsilon_{\rm p} A^2_3 - \frac{{\rm i}}{2}
\epsilon_A \Omega^2_3  & 0 \\
 -{\rm i} \epsilon_{\rm p} A^2_3 + \frac{{\rm i}}{2}
\epsilon_A \Omega^2_3 &
A^2_3 \epsilon_{\rm p} \Omega_2
\tau_{\rm p} & 0 \\ 0 & 0 & 0 \end{array} \right)
\end{equation}
and thus, for $\tau\ge\tau_{\rm tr}$, 
\begin{equation}\label{H18}
\tilde{\rho}^> (\tau_{\rm p} + \tau) = 
U_\pi (\tau) ~\tilde{\rho}^>_{\rm P}~ U^\dagger_\pi (\tau)
\end{equation}
up to the same order in ${\mbox{\boldmath $\epsilon$}}$ as $\tilde{\rho}^> 
(\tau_{\rm p})$
(i.e.\ up to order ${\mbox{\boldmath $\epsilon$}}^2$, unless one 
had started close to $| 2 \rangle$ at the beginning of the probe pulse). 

In a similar way, one finds for the normalized density matrix of the
subensemble without emissions during the probe pulse the density
matrix at time  $\tau_{\rm p}+ \tau$, for $\tau > \tau_{\rm tr}$,  
\begin{equation}\label{H19}
\tilde{\rho}^0 (\tau_{\rm p} + \tau) = 
U_\pi (\tau) ~\tilde{\rho}^0_{\rm P}~  U_\pi^\dagger (\tau)
\end{equation}
with
\begin{equation}\label{H20}
\tilde{\rho}^0_{\rm P} = \left( \begin{array}{ccc}
0 & -{\rm i} \epsilon_{\rm p} & 0 \\
{\rm i} \epsilon_{\rm p} & 1 & 0 \\ 0 & 0 & 0
\end{array} \right)~,
\end{equation}
again up to the same order in ${\mbox{\boldmath $\epsilon$}}$ as 
$\tilde{\rho}^0 (\tau_{\rm p})$.

These results can be viewed in the following way. At the end of a
probe pulse both the subensemble with and without photon emissions still
have components involving the third level, which subsequently decay
during the transient time. Simultaneously the 
population of level 1 is increased correspondingly. In addition,
during this decay, one also has the action of the $rf$ field which
introduces a change dependent on the length of the transient
decay time one considers. However, from Eqs.~(\ref{H18}) and
(\ref{H19}) it is apparent that for times larger than the transient
time one obtains the correct result also if one projects
onto  the states $\tilde{\rho}^>_{\rm P}$ and
$\tilde{\rho}^0_{\rm P}$ at the end of a probe pulse and then develops with the
$rf$ field only. These projection states can be considered virtual in
the sense that they are really never quite realized in the actual
time development of the atom. But it is at least {\em formally} 
consistent to
say that a probe pulse acts as an effective measurement and projects
onto $\tilde{\rho}^>_{\rm P}$ or  $\tilde{\rho}^0_{\rm P}$.

If one neglects all $\epsilon$ terms the states $\tilde{\rho}^>_{\rm P}$
and  $\tilde{\rho}^0_{\rm P}$ become $| 1 \rangle
\langle 1 |$ and $| 2 \rangle \langle 2 |$, respectively, as for an
ideal measurement to which the projection postulate can be
applied. The $\epsilon$ terms can thus be viewed as corrections to the
projection postulate. In the case $\epsilon_A \ll 1$ $(\Omega_2 \ll A_3)$ 
the second term in the 1-2 component of $\tilde{\rho}^>_{\rm P}$
can be neglected, 
and $\tilde{\rho}^>_{\rm P}$ then reduces to Eq.~(61) of Ref. \cite{BeHe}.

\vspace*{0.5cm}             

\noindent {\bf 3. Applications to the Quantum Zeno effect: 
Corrections to the projection-postulate results}

\vspace*{0.5cm}

As shown above a probe pulse, regarded as a measurement pulse, models
very closely  an ideal measurement to which the projection postulate
can be applied. As pointed out in Ref. \cite{BeHe}, for repeated probe
pulses the small differences may add up, however, 
and the net result for an ensemble of atoms was calculated there
in the case $\Omega_3 \ll A_3$. Here this will now be calculated for
an ensemble of atoms
without this restriction for $n$ probe pulses during a $\pi$ pulse of
length $T_\pi = \pi/\Omega_2$ (cf. Fig. 2). The time between two pulses is
denoted by $\Delta T$. It will be assumed that $\Delta T \gg A_3^{-1}$, 
in fact $\Delta T $ larger than the transient decay time $\tau_{\rm
tr}$ will do.

At the end of the $k$-th probe pulse the total ensemble of atoms
consists of two subensembles, one with and the other without
emissions
during the $k$-th  pulse. Hence, by Eqs.~(\ref{H11g}) and (\ref{i17a}),
the total density matrix $\rho(t)$ at time $t = k\cdot 
(\Delta T + \tau_{\rm p})$ is
of the form
\begin{equation}\label{H21}
\rho (t) = \alpha (k) \tilde{\rho}^> + \beta (k) \tilde{\rho}^0 +
{\cal O}({\mbox{\boldmath $\epsilon$}}^2) 
\end{equation}
with $\alpha + \beta =1$. We will now determine $\alpha (k)$
and $\beta (k)$. 

We consider an ensemble with
density matrix $\tilde{\rho}^>$ at the end of a probe pulse, 
develop for the 
time $\Delta T$ without the probe pulse and then again switch on the next
probe pulse. We define $p$ as 
the probability of finding no photons
during this pulse. Similarly, $q$ is defined as 
the no-photon probability for
starting with $\tilde{\rho}^0$ instead of $\tilde{\rho}^>$. We will
calculate $p$ and $q$ explicitly further below.

Now, if the density matrix is given by Eq.~(\ref{H21}) at the end of
the $k$-th  probe pulse, then at the end of the next pulse it is given
by 
\begin{equation}\label{H22}
(1 - p) \alpha (k) \tilde{\rho}^> + p \alpha(k) \tilde{\rho}^0 + (1 -
q) \beta (k) \tilde{\rho}^> + q \beta (k) \tilde{\rho}^0~.
\end{equation}
Hence $\beta$ satisfies the recursion relation
\begin{equation}\label{H23}
\beta (k + 1) = p \alpha(k) + q \beta(k)~.
\end{equation}
Using $\alpha = 1 - \beta$ the solution is seen to be 
\begin{equation}\label{H24}
\beta (k) = p~~\frac{1 - (q-p)^{k-1}}{1 - (q-p)}~ + (q-p)^{k-1} \beta(1)
\end{equation}
for $k > 1$. We determine $\beta (1)$ for the case that at $t = 0$
all atoms are prepared in the ground state. At the beginning of the
first probe pulse the state is then $U_\pi (\Delta T) |1 \rangle$,
and $\beta (1)$ is the probability for no photon emission until the
end of the first pulse. With Eqs.~(\ref{H12}) and (\ref{i16}) and the
abbreviations 
\begin{equation}\label{H25}
c \equiv c(\Delta T) \equiv \cos (\Omega_2 \Delta T)~,
 ~s \equiv s(\Delta T) \equiv \sin (\Omega_2 \Delta T)
\end{equation}
one easily obtains
\begin{equation}\label{H26}
\beta (1) =~\frac{1}{2} (1-c) + \epsilon_{\rm p} s -
 ~\frac{1}{2}~\pi~\frac{\tau_{\rm p}}{T_\pi}~(1-c) \epsilon_{\rm p} + 
{\cal O}({\mbox{\boldmath $\epsilon$}}^2)~.
\end{equation}

In the determination of the no-photon probabilities $p$ and $q$ the
relevant density matrices at the beginning of the next probe
probe pulse are
$U_\pi(\Delta T)\,\tilde\rho^>_{\rm P}\,U_\pi(\Delta T)^\dagger$ and
$U_\pi(\Delta T)\,\tilde\rho^0_{\rm P}\,U_\pi(\Delta T)^\dagger$.
Using Eq.~(\ref{i16})
one obtains, with $\epsilon_A = \Omega_2/A_3$,
\begin{eqnarray}\label{27}
p & = & \frac{1}{2}~(1-c) + \epsilon_{\rm p} \left\{ 2 s~\frac{A^2_3 +
  \Omega^2_3}{A^2_3 + 2 \Omega^2_3}~+~\frac{1}{2}~\Omega_2 \tau_{\rm p}
c~\frac{3 A^2_3 + 2 \Omega^2_3}{A^2_3 + 2 \Omega^2_3}
-~\frac{1}{2}~ \Omega_2 \tau_{\rm p} \right\} \nonumber\\
& &-\frac{1}{2} s~ \frac{\Omega^2_3}{A^2_3 + 2 \Omega^2_3} 
\epsilon_A +{\cal O}({\mbox{\boldmath $\epsilon$}}^2) \nonumber\\
q & = & ~\frac{1}{2} (1 + c) - \epsilon_{\rm p} \left\{ 2 s
+~\frac{1}{2}~\Omega_2 \tau_{\rm p}(1 + c)\right\} + 
{\cal O}({\mbox{\boldmath $\epsilon$}}^2)~. 
\end{eqnarray}

One can now insert $\beta(k)$ of Eq.~(\ref{H24}) and $\alpha(k) = 1 -
\beta(k)$ with $k = n$ into Eq.~(\ref{H21}) to obtain the density
matrix $\rho(t = T_\pi)$ at the end of the $\pi$ pulse, after $n$
probe pulses. Expanding in terms of ${\bf \epsilon}$ one obtains for
the population of level 2
\begin{eqnarray}\label{H28}
\rho_{22}(T_\pi) &=& \frac{1}{2}~(1 - c^n) + \epsilon_{\rm p} \left\{ s
c^{n-1}~ \frac{(2 n-1) A^2_3 + 3 n \Omega^2_3}{A^2_3 + 2
  \Omega^2_3} \right. \nonumber\\
& & \left. + \pi~\frac{\tau_{\rm p}}{T_\pi}~n c^n~\frac{A^2_3 +
\Omega^2_3}{A^2_3 +
  2 \Omega^2_3}~-~\frac{1-c^n}{1 - c}~ \left( s +
\pi~\frac{\tau_{\rm p}}{T_\pi} \right) ~\frac{\Omega^2_3}{A^2_3 + 2
  \Omega^2_3} \right\} \nonumber\\
& & -~\frac{1}{4}~ \epsilon_A s~\frac{\Omega^2_3}{A^2_3 + 2
  \Omega^2_3}~ \left\{ \frac{1 - c^{n-1}}{1 - c}~+ (n - 1) c^{n-1} 
\right\} +{\cal O}({\mbox{\boldmath $\epsilon$}}^2)
\end{eqnarray}
where $c(\Delta T)$ and $s(\Delta T)$ are given by Eq.~(\ref{H25}),
$\epsilon_{\rm p} = A_3 \Omega_2 / \Omega^2_3 \ll 1$ and $\epsilon_A
=  \Omega_2/A_3 \ll 1$. 

This ``quantum jump'' result contains additional $\epsilon$ terms when
compared  with the special case in Eq.~(76) of Ref. \cite{BeHe} for
$\Omega^2_3 \ll A^2_3$, while up to zeroth order it is the same. The
zeroth order gives the approximation
\begin{equation}\label{H29}
\rho_{22}(T_\pi) \cong~\frac{1}{2}~ \left\{ 1 - \cos^n \left(
\frac{1}{n}~-~\frac{\tau_{\rm p}}{T_\pi} \right) \right \}
\end{equation}
while for $n$ {\em instantaneous} ideal measurements the projection
postulate would yield 
\begin{equation}\label{H30}
\frac{1}{2}~\left\{ 1 - \cos^n~\frac{\pi}{n} \right\} ~.
\end{equation}
Thus the zeroth order result in Eq.~(\ref{H29}) can be viewed as the
result of the projection postulate with the finite duration of the
measurement pulse taken into account. The intuitive reason for this
has been discussed in some detail in Ref. \cite{BeHe}.

\begin{table}
\begin{center}
\noindent \begin{tabular}{rccccc}
\hline
\noindent & $~~~~~~~\;$Project & \hspace{-1.45cm}ion Postulate  \\ 
\noindent $n$
& $\Delta T = T_\pi/n$
& $\Delta T = T_\pi/n-\tau_{\rm p}$ 
& Quantum Jump
& Bloch eq. 
& Observed \cite{Wine} \\
\hline
1 & 1.00000 & 0.99978 & 0.99978 & 0.99978 & 0.995 \\
2 & 0.50000 & 0.49957 & 0.49960 & 0.49960 & 0.500 \\
4 & 0.37500 & 0.35985 & 0.36062 & 0.36056 & 0.335 \\
8 & 0.23460 & 0.20857 & 0.20998 & 0.20993 & 0.194 \\
16 & 0.13343 & 0.10029 & 0.10215 & 0.10212 & 0.103 \\
32 & 0.07156 & 0.03642 & 0.03841 & 0.03840 & 0.013 \\
64 & 0.00371 & 0.00613 & 0.00789 & 0.00789 & \llap{$-$}0.006 \\
\hline
\end{tabular}
\parbox{15cm}{\caption { Predicted and observed population 
of level 2 at the end of the $\pi$ pulse for $n$ probe pulses of 
length $\tau_{\rm p}$.}}
\end{center}
\end{table}  

In the following tables we have evaluated the expressions for
$\rho_{22} (T_\pi)$ of Eq.~(\ref{H29}), i.e.\ the (modified)
projection-postulate result, and of Eq.~(\ref{H28}), the quantum jump
result, as well as a numerical solution of the corresponding Bloch
equations of Eq.~(\ref{2.16e}) of the Appendix. 

Table 1 is calculated for the parameters of Ref. \cite{Wine},
i.e.\ $T_\pi =0.256$s, $A_3 = 1.2 \cdot 10^8$/s, $\Omega_3 = 1.9 \cdot
10^6 /$s, and $\tau_{\rm p} = 2.4 \cdot 10^{-3}$s; here $\Omega^2_3 \ll
A^2_3$. The column with the quantum-jump results coincides with the
corresponding column in Table 2 of Ref. \cite{BeHe}. This shows that
the additional $\epsilon$ terms in Eq.~(\ref{H28}) compared to
Eq.~(76) of Ref. \cite{BeHe} are indeed negligible in this case. The
agreement between the quantum-jump results and Bloch equations is
excellent, while the column with the (modified) projection-postulate
results show the small, but still markedly noticeable, relevance of
the $\epsilon$ terms in the quantum-jump expression of
Eq.~(\ref{H28}). In this case the probe or measurement pulses give a
very good, but not perfect, realization of ideal measurements.

\begin{table}
\begin{center}
\noindent \begin{tabular}{rccc}
\hline
\noindent & Projection Postulate  \\ 
\noindent $n$
& $\Delta T = T_\pi/n-\tau_{\rm p}$ 
& Quantum Jump
& Bloch eq. \\
\hline
1 & 0.99978 & 0.99978 & 0.99978 \\
2 & 0.49957 & 0.49957 & 0.49956 \\
4 & 0.35985 & 0.35985 & 0.35979 \\
8 & 0.20857 & 0.20858 & 0.20853 \\
16 & 0.10029 & 0.10030 & 0.10027 \\
32 & 0.03642 & 0.03642 & 0.03641 \\
64 & 0.00613 & 0.00613 & 0.00613 \\
\hline
\end{tabular}
\parbox{15cm}{\caption { Predicted population of level 2 at the end 
of the $\pi$ pulse for the parameters of the experiment, but with 
$\Omega_3=A_3/2$. }}
\end{center}
\end{table}  

A much better realization of ideal measurements can be obtained by
choosing $\Omega_3$ much larger, e.g. of the order of $A_3$. Then one
has a strong pumping between levels 1 and 3, with many photon
emissions, and the last photon during a probe pulse will therefore be
emitted shortly before its end. After this the atom is in the ground
state, and the $rf$ field has little time to build up a correlation
between levels 1 and 2. On the other hand, the population of level 3
at the end of a probe pulse grows with $\Omega^2_3$, and during the
transient decay time after the end of a probe this leads to a
build-up of the 1-2 correlation, but with the opposite sign, as seen
 from $\tilde{\rho}_{\rm P}^>$ in Eq.~(\ref{H17}). The net result is that the
probe pulse projects the subensemble with photon emissions onto a
state much closer to $|1 \rangle \langle 1|$ than for small
$\Omega_3$ \footnote{%
Note that $\epsilon_{\rm p} = \Omega_2 A_3/\Omega_3^2$ decreases 
quadratically 
with $\Omega_3$. Hence the off-diagonal elements in $\rho^>_{\rm P}$ and 
$\rho^0_{\rm P}$ become extremely small when $\Omega_3$ is increased to the 
order of $A_3$.}. 
For large $\Omega_3$ one therefore expects
the projection-postulate results to be much closer to the quantum jump
or Bloch equations results. This is indeed borne out in Table 2 for
$\Omega_3 = A_3/2$, the other parameters being as in Table 1. Now the
projection postulate results (with finite pulse duration) in the first
column agree with the other ones much better, with the difference
starting in the fifth decimal.

\vspace*{0.5cm}

\noindent {\bf 4. Discussion}

\vspace*{0.5cm}

In this paper we have investigated in detail to what extent a short
pulse of a probe laser, which pumps the ground state of an atom 
to a third auxiliary level, can be regarded as a
measurement of the states of the two-level system. In contrast to our
previous paper \cite{BeHe} we have now allowed an arbitrary strength of
the probe pulse. We have shown that in this general case the
projection-postulate result is modified by additional small correction
terms. But we have also shown that for a strong probe pulse,
e.g. $\Omega_3$ in the order of $A_3$, the probe pulse acts in a way
much closer to the ideal projection postulate. This is evident from
the ``virtual'' projection matrices $\tilde{\rho}^>_{\rm P}$ and
$\tilde{\rho}^0_{\rm P}$ of Eqs.~(\ref{H18}) and (\ref{H19}) which 
for large $\Omega_3$ are
extremely close to $|1 \rangle \langle 1|$ and $|2 \rangle \langle
2|$, respectively. Indeed, the off-diagonal
elements, which are already small for weak probe pulses, become orders
of magnitude smaller for strong probe pulses.

The slow-down of the time development of a state under repeated probe
pulses (``measurements'') is apparent in Tables 1 and 2. For strong
probe pulses this is much closer to the projection-postulate
predictions (with finite pulse duration taken into account). But an
actual freezing of the state, as predicted by the projection postulate
for instantaneous ideal measurement for $\Delta t \rightarrow 0$, 
is of course still not obtainable
by these probe pulses. If one decreases the time $\Delta T$ between
the pulse to the order of the transient time $\tau_{\rm tr}$ 
or even to $A_3^{-1}$, then
the third level does not decay completely to level 1, and the probe
pulse can no longer be regarded as a measurement pulse.

\vspace*{0.5cm}

\noindent{\bf Appendix: The quantum jump approach in quantum optics}

\vspace*{0.5cm}

We briefly summarize the quantum jump approach used in this paper. 
The  quantum jump approach \cite{HeWi,Wi,He,HeSo}, quantum trajectories 
\cite{QT} and the Monte-Carlo wave function approach 
\cite{MC} are essentially equivalent. It describes
a radiating atom between photon detections by a 
reduced (or conditional)
time evolution operator giving the time development under the
condition that no photon has been detected \cite{HeWi}. After a photon 
detection one has to reset the atom to the reset state (``jump''),
with ensuing reduced time development, and so on. The general reset 
states have been determined in Ref. \cite{He}; cf. also Ref. \cite{HeSo}.
For a driven  system with many emissions one then obtains a
stochastic path, also called a quantum trajectory \cite{QT}.
For a V system as considered in this paper the reset state after an 
emission is  the ground state. The reduced time development together with 
the reset states provide a complete stochastic description 
of the time development of the atom \cite{He,HeSo}. 
Starting with this description 
one can then derive the Bloch equations describing an ensemble of radiating
atoms \cite{HeWi,He}. 

We consider the V system depicted in Fig.~1. In this system
the upper levels 2 and 3 couple to a common ground level 1, with Einstein 
coefficient $A_3$ (in this paper level 2 is taken as stable).
We assume here that
$\omega_{32}~\equiv~\omega_{3} - \omega_{2} $ is in the optical
range. For simplicity we consider zero detunings
of the driving fields, whose (real) 
Rabi frequencies are denoted by $\Omega_2$
and $\Omega_3$, respectively. In the interaction picture with respect
to the free atomic Hamiltonian $H^{\rm A}_0$,  the reduced
Hamiltonian $H^I_{\rm red}$ is given 
by \cite{BeHe,HeWi,HeSo}
\begin{equation}
H^I_{\rm red}/\hbar~=  \frac{1}{2}~\left( \begin{array}{ccc}
0 & \Omega_2  & \Omega_3 \nonumber\\
\Omega_2  &  0 & 0 \\
\Omega_3 & 0 &-{\rm i}\,A_3 \nonumber
\end{array}\right)
 ~\equiv -{\rm i}M \,.\label{2.12}
\end{equation}
where the atomic operator $M$ is defined by the l.h.s. and where we
have used matrix notation with respect to the atomic basis
$|1\rangle$, $|2\rangle$, $|3\rangle$. The time development of an atom 
between emissions is then given by 
\begin{equation}
U^I_{\rm red}(\tau) = {\rm e}^{-{\rm i}H^I_{\rm red}\tau/\hbar}
= {\rm e}^{-M\tau}~. \label{2.15}
\end{equation}
The no-photon probability until time $\tau$ is then, for initial state
$|\psi \rangle$, 
\begin{equation} \label{2.16}
P_0(\tau;|\psi \rangle) = \| {\rm e}^{-M\tau} | \psi \rangle \|^2
\end{equation}
or, more generally for an initial density matrix $\rho$,
\begin{equation}\label{2.16a}
P_0(\tau;\rho) = {\rm tr} \left\{ {\rm e}^{-M\tau} \rho
{\rm e}^{-M^{\dagger}\tau}
\right\} ~. \nonumber
\end{equation}
The probability for the first photon to be emitted in $(\tau,\tau+ d\tau)$
equals $P_0(\tau;\rho) - P_0(\tau +d\tau;\rho) 
\equiv w_1 (\tau;\rho)d\tau$, where
\begin{eqnarray} \label{2.17}
w_1( \tau ; |\rho) = - \frac{{\rm d}}{{\rm d}\tau}~ P_0(\tau;\rho)
\end{eqnarray}
is the probability density for the first photon\footnote{%
Depending on the parameters  there may be a finite 
probability that no photon is emitted at all. Therefore  this probability 
density need not be normalized to 1.}. 
For small upper level separation nonzero off-diagonal generalized damping terms
may appear which lead to interesting coherence effects 
\cite{HePl1,HePl2,HePl3,HePl4}. For general $n$-level systems the
reduced Hamiltonian is derived in Ref. \cite{He,HeSo}.

The reduced time development is not unitary. The reason is that
it does not describe the time evolution of the whole ensemble but
that of the sub\-ensemble with no photons. The size of this sub\-ensemble
is decreasing in time since an atom for which a photon has been detected
leaves the sub\-ensemble, and this is reflected in the decrease of the norm 
squared in Eq.~(\ref{2.16}). The above probability density
determines the (random) time for the first photon. After that the atom
is reset, for a V system to the ground state, $|1\rangle$. 
The next emission time is then determined by $w_1(\tau;|1\rangle)$, and so 
on. In this way one obtains a quantum trajectory.

 From this description of single systems
one can recover the usual Bloch equations of the complete ensemble as
follows \cite{He}. The density matrix $\rho(\tau)$ of the ensemble is a
sum of two terms, $\rho^>$ and $\rho^0$, corresponding to a
subensemble of atoms with and without photon emissions until time $\tau$,
respectively,
\begin{eqnarray}\label{2.16b}
\rho(\tau) = \rho^0(\tau) + \rho^>(\tau)~.
\end{eqnarray}
 From Eq.~(\ref{2.15}) one has for initial state $\rho$ 
\begin{equation}\label{2.B1}
\rho^0 ( \tau ; \rho )  =  {\rm e}^{- M \tau} ~\rho~{\rm e}^{- M^{\dagger}
\tau}~.
\end{equation}
If $I(\tau';\rho)d\tau'$ denotes the (unconditioned) probability to
find a photon in ($\tau',\tau'+d\tau'$) then, at time $\tau$,
the sub-subensemble of atoms with their last emission in this interval is
described by 
\begin{equation}\label{2.B2}
I(\tau';\rho)d\tau' \rho^0(\tau - \tau';|1\rangle)
\end{equation}
and therefore
\begin{equation} \label{2.B3}
\rho^> (\tau;\rho) = ~\int^\tau_0~{\rm d}\tau'~I(\tau' ; \rho )
\rho^0 (\tau - \tau';|1 \rangle)~.
\end{equation}
Thus one has
\begin{equation}\label{A}
\rho(\tau) = {\rm e}^{- M \tau} ~\rho~{\rm e}^{- M^{\dagger} \tau}
        + \int^\tau_0~{\rm d}\tau'~I(\tau' ; \rho )
\rho^0 (\tau - \tau';|1 \rangle)~.
\end{equation}
Differentiation gives 
\begin{equation}\label{2.B4}
\dot\rho(\tau;\rho) = \dot\rho^0 ( \tau ; \rho ) + 
I(\tau;\rho)|1\rangle \langle1| + ~\int^{\tau}_0~{\rm d}
\tau'~I(\tau' ; \rho ) \dot\rho^0 (\tau - \tau';|1\rangle)~.
\end{equation}
Taking the trace and using $\mathop{\rm tr}\rho(\tau) \equiv 1$ gives
\begin{eqnarray}\label{2.16d}
I(\tau ; \rho) = A_3 \rho_{33}(\tau;\rho)
\end{eqnarray}
and thus Eq.~(\ref{A}) becomes in the present situation
\begin{eqnarray}\label{2.16e}
\rho ( \tau ; \rho )  =  {\rm e}^{- M \tau} \rho {\rm e}^{- M^{\dagger} \tau}
+ \int^{\tau}_0~{\rm d}\tau'~A_3 \rho_{33}(\tau';\rho) \rho^0
(\tau - \tau';|1\rangle)~.
\end{eqnarray}
 From Eq.~(\ref{2.B1}) one obtains $\dot\rho^0$, 
and inserting this into Eq.~(\ref{2.B4}) gives 
\begin{equation} \label{2.B5}
\dot\rho(\tau) = -\frac{{\rm i}}{\hbar}[H^I_{\rm red}\rho(\tau) - 
\rho(\tau)
H^{I\dagger}_{\rm red}] + A_3 \rho_{33}(\tau)|1\rangle \langle1|~.
\end{equation}
This is a compact form of the Bloch equations used in Refs.\ %
\cite{Block,Schenzle}. Conversely, from Eq.~(\ref{2.B5}) one can
immediately obtain  the integral equation of Eq.~(\ref{2.16e}).

\newpage

\begin{figure}
\unitlength 0.6cm
\begin{picture}(18,11)
\thicklines
\put(6,9.5) {\line(1,0){3}}
\put(10,3) {\line(1,0){3}}
\put(7.5,1.5) {\line(1,0){4}}
\thinlines
\put(4.5,3.5) {\line(1,0){0.5}}
\put(5.5,3.5) {\line(1,0){0.5}}
\put(5,5.5) {\line(1,0){0.5}}
\put(5,3.5) {\line(0,1){2}}
\put(5.5,3.5) {\line(0,1){2}}
\put(7,9.5) {\vector(1,-4){2}}
\put(7.5,9.5) {\vector(1,-4){2}}
\put(9,1.5) {\vector(-1,4){2}}
\put(10,1.5) {\vector(1,1){1.5}}
\put(11.5,3) {\vector(-1,-1){1.5}}
\put (5.4,9.5){3}
\put (6,4.25){$\Longrightarrow $}
\put (6.75,6){$\Omega_3$}
\put (8.7,6){$A_3$}
\put (4,2.75){probe pulse}
\put (13.25,3){2}
\put (12.75,2){$\Longleftarrow $}
\put (14.25,2){{\it rf} field, $\Omega_2$}
\put (11.75,1.5){1}
\end{picture}

{\caption{ V system with (meta-)stable level 2 and auxiliary level 3
with Einstein coefficient $A_3$. $\Omega_2$ and $\Omega_3$ are the Rabi
frequencies of the $rf$ field and the probe laser,  respectively.}}
\end{figure}
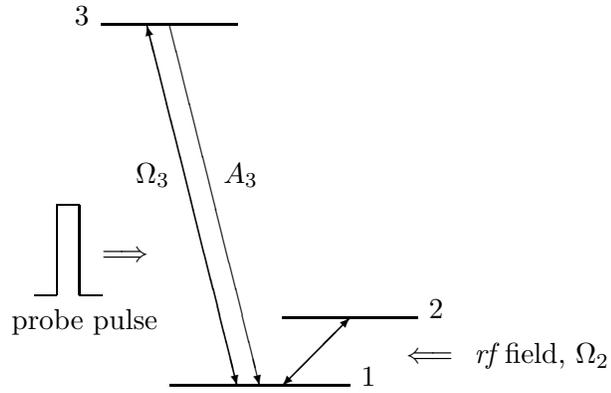

\begin{figure}
\unitlength 0.9cm
\begin{picture}(15.5,6.5)
\thicklines
\put (2,1) {\vector (1,0) {12}}
\put (2,1) {\vector (0,1) {4.5}}
\thinlines
\put (3.5,4.5) {\line (1,0){0.5}}
\put (5.5,4.5) {\line (1,0){0.5}}
\put (7.5,4.5) {\line (1,0){0.5}}
\put (11,4.5) {\line (1,0){0.5}}
\put (2,1.5) {\line (1,0){7}}
\put (10,1.5) {\line (1,0){1.5}}
\put (3.5,1) {\line (0,1){3.5}}
\put (5.5,1) {\line (0,1){3.5}}
\put (7.5,1) {\line (0,1){3.5}}
\put (11,1) {\line (0,1){3.5}}
\put (4,1) {\line (0,1){3.5}}
\put (6,1) {\line (0,1){3.5}}
\put (8,1) {\line (0,1){3.5}}
\put (11.5,1) {\line (0,1){3.5}}
\put (4,3) {\vector (1,0) {1.5}}
\put (5.5,3) {\vector (-1,0) {1.5}}
\put (7,4.25) {\vector (1,0) {0.5}}
\put (8.5,4.25) {\vector (-1,0) {0.5}}
\put(4.5,3.25){$\Delta T$}
\put(8.25,4.5){$\tau_{\rm p}$}
\put(11.75,4.0){probe pulse}
\put(8.35,1.75){$\pi$ pulse}
\put(14.25,0.75){$t$}
\put(9.35,1.5){...}
\put(1.75,0.25){0}
\put(11.35,0.25){$T_{\rm \pi}$}
\put(1,5.2){$\Omega$}
\end{picture}
{\caption{Probe pulses and $\pi$ pulse}}
\end{figure}
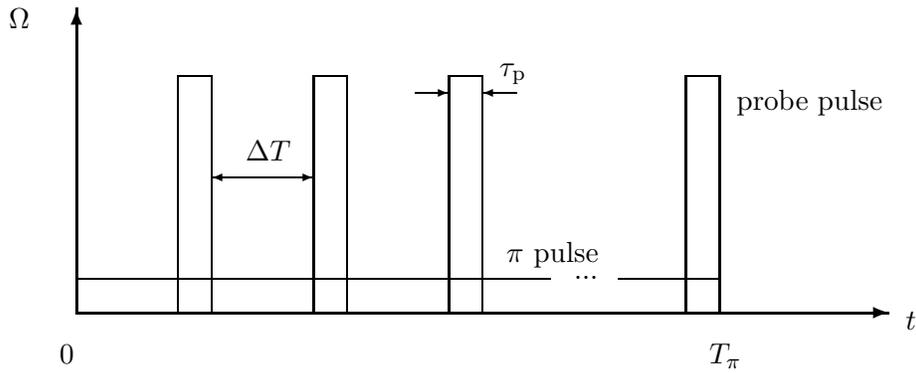

\end{document}